# Strong bulk Dzyaloshinskii-Moriya interaction in composition-uniform centrosymmetric magnetic single layers

Lijun Zhu,[1*] David Lujan,[2] Xiaoqin Li[2]

1. State Key Laboratory of Superlattices and Microstructures, Institute of Semiconductors, Chinese Academy of Sciences, Beijing 100083, China

2. Department of Physics, Center for Complex Quantum Systems, The University of Texas at Austin, Austin, Texas 78712, USA

*ljzhu@semi.ac.cn

**Abstract:** Dzyaloshinskii-Moriya interaction (DMI) is the key ingredient of chiral spintronic phenomena and the emerging technologies based on such phenomena. A nonzero DMI usually occurs at magnetic interfaces or within non-centrosymmetric single crystals. Here, we report the observation of a strong unexpected DMI within a centrosymmetric polycrystalline ferromagnet that has neither a crystal inversion symmetry breaking nor a composition gradient. This DMI is a bulk effect, increases with the thickness of the magnetic layer, and is insensitive to the symmetry of the interfaces or the neighboring materials. We observe a total DMI strength that is a factor of >2 greater than the highest interfacial DMI in the literature. This DMI most likely arises from the strong spin-orbit coupling, strong orbital hybridization, and a "hidden" long-range asymmetry in the material. Our discovery of the strong unconventional bulk DMI in centrosymmetric, composition-uniform magnetic single layers provide fundamental building blocks for the emerging field of spintronics and will stimulate the exploitation of unconventional spin-orbit phenomena in a wide range of materials.

Dzyaloshinskii-Moriya interaction (DMI) is a short-range anti-symmetric exchange interaction that promotes chiral spin textures [1-3]. The DMI has become a foundational aspect of spintronics because it can stabilize chiral spin textures (e.g. magnetic skyrmions [1] or chiral domain walls [2,3]) in emerging memory and computing technologies [4,5] and because it affects the magnetic uniformity, speed [6,7], reliability [8], stability [9], and power [10] of magnetic tunnel junctions. DMI usually occur at magnetic interfaces or in non-centrosymmetric magnetic crystals with broken inversion symmetry [11-15]. In the bulk of compositionally-uniform centrosymmetric films, DMI and other symmetry-breaking effects are "naturally" expected to be absent [16].

However, recent discovery of bulk spin-orbit torques in composition-uniform disordered magnetic alloys (e.g. CoPt, FeTb, and CoTb)[17-19] does imply that there is a new type of symmetry-breaking that is "hidden" but sufficient to result in sizeable unconventional symmetry-breaking effects. In this letter, we report the first observation of a very strong bulk DMI within compositionally uniform single layers of A1-phased CoPt, a representative centrosymmetric ferromagnetic material that has strong spin-orbit coupling (SOC)[20], strong orbital hybridization [20], and strong Brillouin light scattering (BLS) response. The total DMI strength $D_s$ for a 24 nm CoPt reaches -0.34 µerg/cm, which is more than a factor of 2 greater than the highest reported interfacial DMI. We find that this DMI is a bulk effect, with the strength increasing monotonically with the thickness ($t_{FM}$) of the ferromagnetic layer, and is insensitive to the symmetry of the interfaces or the types of the neighboring materials.

We sputter-deposited composition-uniform CoPt (=$Co_{0.5}Pt_{0.5}$) films with different thicknesses ($t_{CoPt}$) of 4, 8, 12, 16, and 24 nm at room temperature on thermally oxidized Si substrates with an argon pressure of 2 mTorr and a base pressure of ~$10^{-9}$ Torr. The deposition rates for Co and Pt are ~0.011 nm/s. Each sample was capped by a MgO 2 nm/Ta 1.5 nm bilayer. Scanning transmission microscopy (STEM) and electron-energy loss spectrum (EELS) measurements indicate good composition homogeneity and minimal oxidation of the CoPt in these samples (Fig. S1, the Supplementary Material [21]). As we discuss below, these CoPt layers have polycrystalline A1 structure, in-plane magnetic anisotropy, and good magnetic homogeneity.

The DMI of the CoPt layers is determined by measuring DMI-induced frequency difference ($\Delta f_{DMI}$) between counter-propagating Damon-Eshbach spin waves using BLS [22-29]. Figure 1(a) shows the geometry of the BLS measurements. The laser wavelength ($\lambda$) is 532 nm. The light incident angle ($\theta$) with respect to the film normal was varied from 0° to 60° to tune the magnon wave-vector ($|k| = 4\pi \sin\theta/\lambda$). A magnetic field ($H$) of ± 2 kOe was applied along the $x$ direction to align the magnetization of the CoPt layers. The Stokes (anti-Stokes) peaks in BLS spectra (Fig. 1(b)) correspond to the creation (annihilation) of magnons with -$k$ ($k$), while the total in-plane momentum is conserved during the light scattering process. In Fig. 1(c), we plot $\Delta f_{DMI}$ as a function of $|k|$ measured from CoPt films with different thicknesses $t_{CoPt}$. Here, $\Delta f_{DMI}$ is the frequency ($f$) difference between the Stokes and anti-Stokes peaks (i.e. $\Delta f_{DMI} = f(+k) - f(-k)$) and it is an averaged value taken by reversing the direction of the applied field $H = \pm 2$ kOe (see [29] for more details). As we discuss below, the variation of $\Delta f_{DMI}$ with $k$ is unrelated to any magnetic asymmetry between the two interfaces of the CoPt. The linear relation between $\Delta f_{DMI}$ and $k$ for each thickness agrees with the expected relation [22-29]

$$\Delta f_{DMI} \approx (2\gamma/\pi\mu_0 M_s)Dk, \tag{1}$$

where $\gamma$ is the gyromagnetic ratio, and $D$ is the volume-averaged DMI constant [22]. As derived theoretically, Eq. (1) holds not only for the interface-type DMI [30,31] but also for bulk-type DMI, e.g., within non-centrosymmetric magnetic crystals or ferromagnets (FMs) that have a $C_{nv}$ or uniaxial crystallographic symmetry (Table 1 in Ref. [32]). Theories have also shown that, in a broad range of thicknesses and wave-vector values, typically when $kt_{FM} < 1.5$ (or the dipole field parameter $P \equiv 1-[1-\exp(-|k|t_{FM})]/$ $|k|t_{FM} = 0.5$ in Ref. [30]), the DE spin wave dispersion and Eq. (1) can be obtained accurately by approximating a magnetic layer with interfacial DMI as an uniform bulk magnet with averaged dynamic variables (e.g., dipole field and the DMI field) over the thickness. For this reason, regardless of the nature of the DMI (either interface- or bulk-type), the total strength of the DMI for a magnetic layer can be estimated as [22,30-32]

$$D_s = D t_{FM}, \qquad (2)$$

Therefore, both Eq. (1) and Eq. (2) should work well in our present experiment in which $k t_{FM} \lesssim 0.48$ (or $P \lesssim 0.2$, Fig. S2 in the Supplementary Materials [21]). Equation (2) also provides the scientific basis for the inverse proportionality of the measured $D$ to the magnetic layer thickness [25-27,30,31] assuming no degradation of the sample quality during the thickness variation process.

As plotted in Fig. 1(d), $D$ increases gradually from $0.055 \pm 0.024$ erg/cm$^2$ at $t_{CoPt} = 4$ nm to $0.140 \pm 0.010$ erg/cm$^2$ at $t_{CoPt} = 16$ nm and then saturates as $t_{CoPt}$ further increases. The total strength of the DMI, $D_s$, increases from $-0.022 \pm 0.010$ μerg/cm$^2$ at 4 nm to $-0.338 \pm 0.007$ μerg/cm$^2$ at $t_{CoPt} =24$ nm (Fig. 1(e)). We first note that, in analogue to that of Pt/Co interface, the observed DMI of the CoPt has a negative sign and lies in the film plane. Moreover, it is very strong compared to the reported interfacial DMI of heavy metal/ferromagnet (HM/FM) interfaces. Despite of the very large CoPt thicknesses, the DMI yields a $D$ value of ≈0.14 erg/cm$^2$ for the 16 nm and 24 nm films, which is two time greater than that of Ta/CoFeB and W/CoFeB with a CoFeB thickness of only 1 nm [29]. Considering the inverse proportionality to the FM thickness, $D$ is a not a fair indicator for the total strength of a given DMI when comparing FM samples with different thickness. In contrast, $D_s$ is a constant and always represents the total strength of the DMI regardless of the FM thickness. Strikingly, the $D_s$ value for the 24 nm CoPt (≈ -0.34 μerg/cm) is more than a factor of 2 greater than the highest reported interfacial DMI: the average of -0.14 μerg/cm for Au$_{0.85}$Pt$_{0.15}$ 4 nm/Co 3.6 nm [33], -0.15 μerg/cm for Pt 4 nm/Fe$_{0.6}$Co$_{0.2}$B$_{0.2}$ 2.6 nm [34], -0.08 μerg/cm for Pt 2-6 nm/Fe$_{0.4}$Co$_{0.4}$B$_{0.2}$ 2 nm [27,35], -0.16 μerg/cm for Pt 4 nm/Co 3.6 nm [33] and Pt 5 nm/Co 1 nm [29], and 0.14 μerg/cm for Ir 7 nm/Co 0.9 nm [36].

More intriguingly, the thickness dependences of $D$ and $D_s$ of the CoPt layers are strong characteristics of a bulk DMI effect because, for an interfacial DMI, $D$ should be inversely proportional to the FM thickness and $D_s$ should remain constant as a function of the FM thickness [25-27]. We also find that the DMI is insensitive to details of the CoPt interfaces, whether interfaces are formed with neighboring insulating materials (SiO$_2$ and MgO) or with metallic layers (Tb). As shown in Figs. 2(a) and 2(b), $D_s$ is ≈ -0.22 μerg/cm$^2$ for the symmetric MgO/CoPt 16 nm/MgO, Tb/CoPt 16 nm/Tb, and also for the asymmetric SiO$_2$/CoPt 16 nm/MgO. Both the unique thickness dependence and the insensitivity to the CoPt interfaces make the observed DMI distinctly different from any interfacial DMI.

The insensitivity of the $D_s$ to the interface details also reaffirms that the measured $D_s$ values are predominantly contributed by the DMI effect and that there is minimal spin wave nonreciprocity from the two interfaces of the CoPt films. This is consistent with the wide consensus that the spin wave frequency shift is neglible at FM/MgO interfaces of sputter-deposited samples [22,23,27,29,33,34,36]. Note that our FM/MgO interfaces are sharp and the FM layers are unoxidized as indicated by the results of STEM [18,33,37], EELS [18], magnetic damping [32], and the thickness dependence of magnetic moment [33].

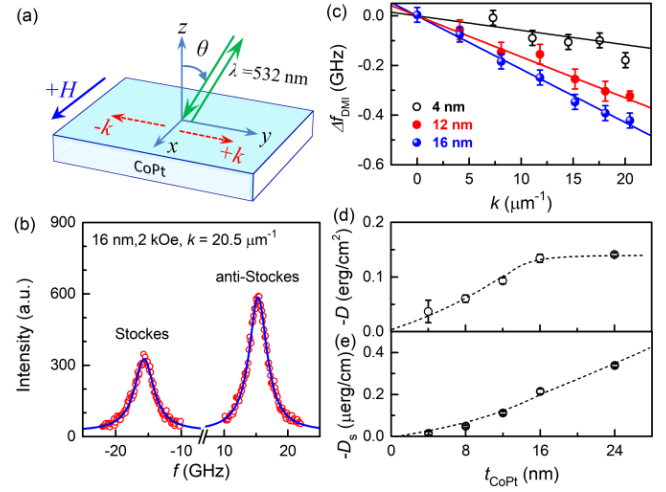

Fig. 1. (a) BLS measurement geometry; (b) BLS spectra ($k$ = 20.5 μm$^{-1}$, $H$ = 2 kOe, $t_{CoPt}$ = 16 nm), (c) $k$ dependence of $\Delta f_{DMI}$ ($t_{CoPt}$= 4, 12, and 16 nm), (d) $-D$, and (e) $-D_s$ for the CoPt layers with different thicknesses. The blue solid curves in (b) represent fits to the Lorentzian function. The solid lines in (c) refer to the best linear fits. The dashed lines in (d) and (e) are to guide the eyes.

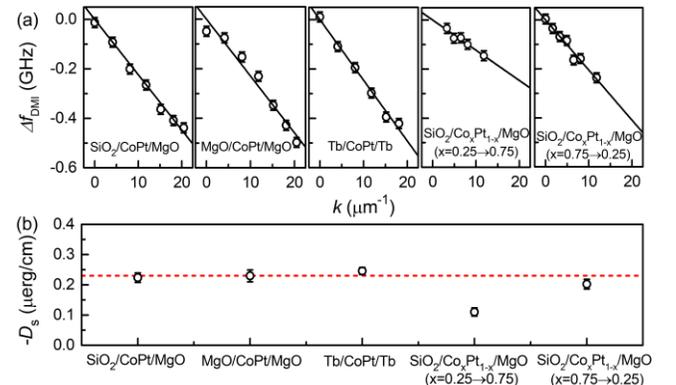

Fig. 2. Effects of interfaces and vertical composition gradient. (a) $k$ dependence of $\Delta f_{DMI}$ and (b) $D_s$ for SiO$_2$/ CoPt 16 nm/MgO 2 nm, MgO 2 nm/ CoPt 16 nm/MgO 2 nm, Tb 2.5 nm/CoPt 16 nm/Tb 2.5 nm, SiO$_2$/Co$_x$Pt$_{1-x}$ ($x$ = 0.25→0.75) 16 nm/MgO, and SiO$_2$/Co$_x$Pt$_{1-x}$ ($x$ = 0.75→0.25) 16 nm/MgO. The solid straight lines in (a) represent the best linear fits. The dashed line in (b) is for the guidance of eyes.

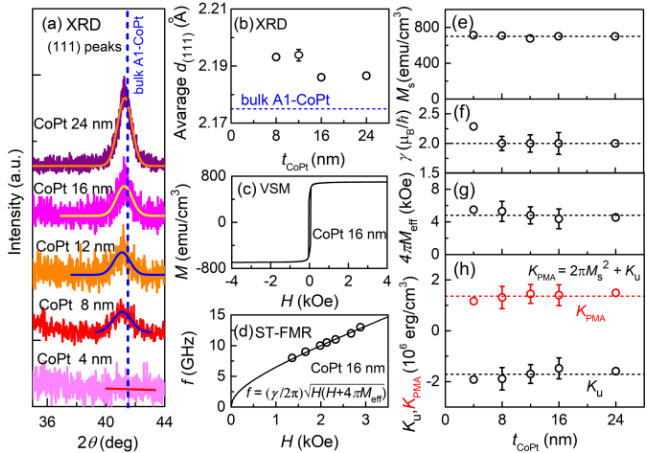

Fig. 3. Structural and magnetic characterizations. (a) $\theta$-$2\theta$ x-ray diffraction patterns and (b) the average spacing of (111) lattice plane, $d_{(111)}$, for the composition-uniform CoPt films. The solid curves in (a) represent the Gaussian fits. The blue dashed lines in (a) and (b) represent the peak position and the spacing of (111) lattice plane for the bulk A1-CoPt [49], respectively. (c) In-plane magnetization for the SiO$_2$/CoPt 16 nm/MgO 2 nm/Ta 1.5 nm measured as a function of in-plane applied magnetic field; (d) Microwave frequency vs ferromagnetic resonance field of the 16 nm CoPt as measured by spin-torque ferromagnetic resonance (ST-FMR). Layer thickness dependences of (e) saturation magnetization, (f) gyromagnetic ratio, (g) $4\pi M_{\text{eff}}$, and (h) uniaxial magnetic anisotropy ($K_{\text{u}}$) and perpendicular magnetic anisotropy ($K_{\text{PMA}}$) for the composition-uniform CoPt samples. The dashed lines in (b), (e), (f), (g), and (h) are to guide the eyes. The solid line in (d) represents the fit to Kittel's Equation. In (b), (d), (e), (f), (g), and (h), some error bars are smaller than the data points.

The occurrence of DMI within the CoPt requires SOC [22,34], orbital hybridization at the Fermi level [33,38–40], and inversion asymmetry. As calculated by theories [20], the CoPt alloys have both strong SOC and orbital hybridization at the Fermi level. However, the mechanism of inversion asymmetry within the CoPt single layers remains a question.

We have found no evidence for a long-range asymmetry in these CoPt films. X-ray diffraction (XRD, Fig. 3(a)) and electron diffraction patterns (Fig. S1 [21]) reveal that these CoPt layers have polycrystalline A1 structure with a preferred (111) orientation. Such structure is globally centrosymmetric and cannot provide a crystal inversion asymmetry necessary for the bulk DMI. The insensitivity of the DMI to the details of the interfaces excludes any important role of interface effects (e.g. oxidization-related interfacial DMI [41–43]) in producing associated inversion asymmetry. These CoPt layers also have good composition homogeneity across the thickness, as revealed by the uniform contrast in scanning transmission electron microscopy image and by the fairly constant EELS intensities of Co and Pt in the depth profile (Fig. S1 [21]). We have also found that a strong artificial vertical composition gradient may even reduce the apparent $D_s$ strength in the CoPt systems. As shown in Figs. 2(a) and 2(b), $D_s$ is -0.110±0.013 μerg/cm$^2$ for a control sample of 16 nm Co$_x$Pt$_{1-x}$ with $x$ continuously varying from 0.25 at the bottom to 0.75 at the top ($x$ = 0.75→0.25), which is a factor of 2 smaller than that of the 16 nm uniform CoPt (-0.224±0.016). Another 16 nm Co$_x$Pt$_{1-x}$ with the reversed composition gradient ($x$ = 0.75→0.25) shows the DMI of the same sign (-0.202±0.016). This indicates that neither the magnitude nor the sign of the DMI is caused directly by a composition gradient or asymmetry in composition-sensitive effects (e.g. magnetic moment density, electronegativity [4], or orbital hybridization [33,38–40]).

Since DMI can be sensitive to strain [44–48], a strain gradient resulting from an accelerating strain relaxation along film normal could potentially break the inversion symmetry for the observed thickness dependence of $D$ and $D_s$ (see Section 3 in the Supplementary Material [21] for a more detailed discussion). However, there is no difference in the lattice constants at the top and bottom regions of the CoPt within the resolution of cross-sectional electron diffraction (Fig. S1 [21]). It is also very challenging to verify whether there was an increasing density of dislocations in the polycrystalline CoPt by transmission electron microscopy because of the overlap of grains with different orientations [18]. $\theta$-$2\theta$ XRD results show that the average out-of-plane lattice plane spacing, $d_{(111)}$, is smaller than the bulk A1-CoPt value [49] of 2.175 Å by 0.9% for $t_{\text{CoPt}} \leq 12$ nm and by 0.5% when $t_{\text{CoPt}}$ is 16 nm and 24 nm (Fig. 3(a) and 3(b)). While this suggests a vertical tensile strain (formed during the sputtering deposition due to thermal and/or substrate stress [50–54]), the XRD results do not clearly indicate a monotonic decrease of $d_{(111)}$ with the CoPt thickness or a strain gradient.

Finally, there is no apparent suggestion of a gradient in the magnetic properties (Figs. 3(c)-3(g))). $M_s$ is estimated by vibrating sample magnetometry (VSM) to be 700 ± 10 emu/cm$^3$ for all these CoPt films (Figs. 3(c) and 3(e)). These CoPt samples also show similar $\gamma$ of ≈ 2 $\mu_B/h$ (Figs. 3(f)), effective demagnetization field ($4\pi M_{\text{eff}}$) of ≈ 4.5 kOe (Figs. 3(g)), uniaxial magnetic anisotropy $K_{\text{u}}$ (= $4\pi M_{\text{eff}} M_s/2$) of ≈ -1.71 ± 0.19 erg/cm$^3$, and perpendicular magnetic anisotropy $K_{\text{PMA}}$ (= $2\pi M_s^2 + K_{\text{u}}$) of ≈ -1.36 ± 0.13 erg/cm$^3$. Here, $\gamma$ and $4\pi M_{\text{eff}}$ are determined by fits of the microwave frequency ($f$) dependence of the ferromagnetic resonance field ($H_r$) to the Kittel's Equation [55] $f = (2\pi/\gamma)\sqrt{H_r(H_r + 4\pi M_{\text{eff}})}$ (Fig. 3(d) and Fig. S5 [21]).

Although there is no theory predicting the novel bulk DMI reported here, our experimental observations imply that, in magnetic materials with a strong SOC and strong orbital hybridization, a bulk DMI can be generated without the need of an obvious long-range asymmetry in crystal structure or magnetism. We speculate that the inversion asymmetry that allows the DMI to occur could be associated with a "hidden" local effect (such as short-range order or local SOC effect). Another example for "hidden" local effects is the bulk perpendicular magnetic anisotropy in thick, amorphous ferrimagnets, e.g. FeTb [19].

In summary, we have demonstrated the observation of a strong bulk DMI within magnetic single layers that have no broken crystal inversion symmetry and composition non-uniformity. This DMI increases with the thickness of the magnetic layer, and is insensitive to the symmetry of the interfaces or the types of the neighboring materials, which are clear bulk features. The bulk DMI can be more than a factor of 2 stronger than the highest conventional interfacial DMI even in composition-uniform ferromagnetic layers. The bulk DMI within the CoPt is attributed to strong SOC, strong orbital hybridization at the Fermi level, and a "hidden" inversion asymmetry.

Such strong unconventional bulk DMI that does not require any crystal inversion asymmetry or composition gradient provides new, simple, integration-friendly, fundamental building blocks of the emerging field of spintronics. The bulk DMI is more effective than interfacial DMI for creating strong chiral effects in applications that prefer a relatively thick magnetic layer. Compared to the conventional interfacial DMI of magnetic superlattices, the bulk DMI can substantially simplify the material deposition process and significant improve the scalability and the stability of magnetic memory and computing technologies that utilize magnetic skyrmions and domain walls. The strong bulk DMI also greatly enriches the material options for in-depth study of the DMI effects on the speed, reliability, stability, and power of nanoscale magnetic tunnel junctions. Our finding of the strong unconventional bulk DMI in centrosymmetric and composition-uniform materials should fundamentally broaden the scope of spintronics and will stimulate future experimental and theoretical exploitation of the effects and spin-orbit technologies in a wide range of material systems. We speculate that the hidden mechanism of inversion symmetry-breaking, if unveiled in the future, may be used to enable and tune the bulk DMI and other unconventional SOC phenomena (e.g. bulk spin-orbit torques [17-19], bulk exchange bias, etc. ).

We thank Kirill Belashchenko for fruitful discussions about the strain profile. Lijun Zhu acknowledges the Strategic Priority Research Program of the Chinese Academy of Sciences (XDB44000000) and the start-up funding support from Institute of Semiconductors, Chinese Academy of Sciences. The DMI measurement performed at UT-Austin was primarily supported by the Center for Dynamics and Control of Materials: an NSF MRSEC under Cooperative Agreement No. DMR1720595.

Supplementary Materials for

# Strong bulk Dzyaloshinskii-Moriya interaction in centrosymmetric magnetic single layers

Lijun Zhu,[1,*] David Lujan,[2] Xiaoqin Li[2]

*1. State Key Laboratory of Superlattices and Microstructures, Institute of Semiconductors, Chinese Academy of Sciences, Beijing 100083, China*

*2. Department of Physics, Center for Complex Quantum Systems, The University of Texas at Austin, Austin, Texas 78712, USA*

*\*ljzhu@semi.ac.cn*

## Section 1. Scanning transmission electron microscopy results

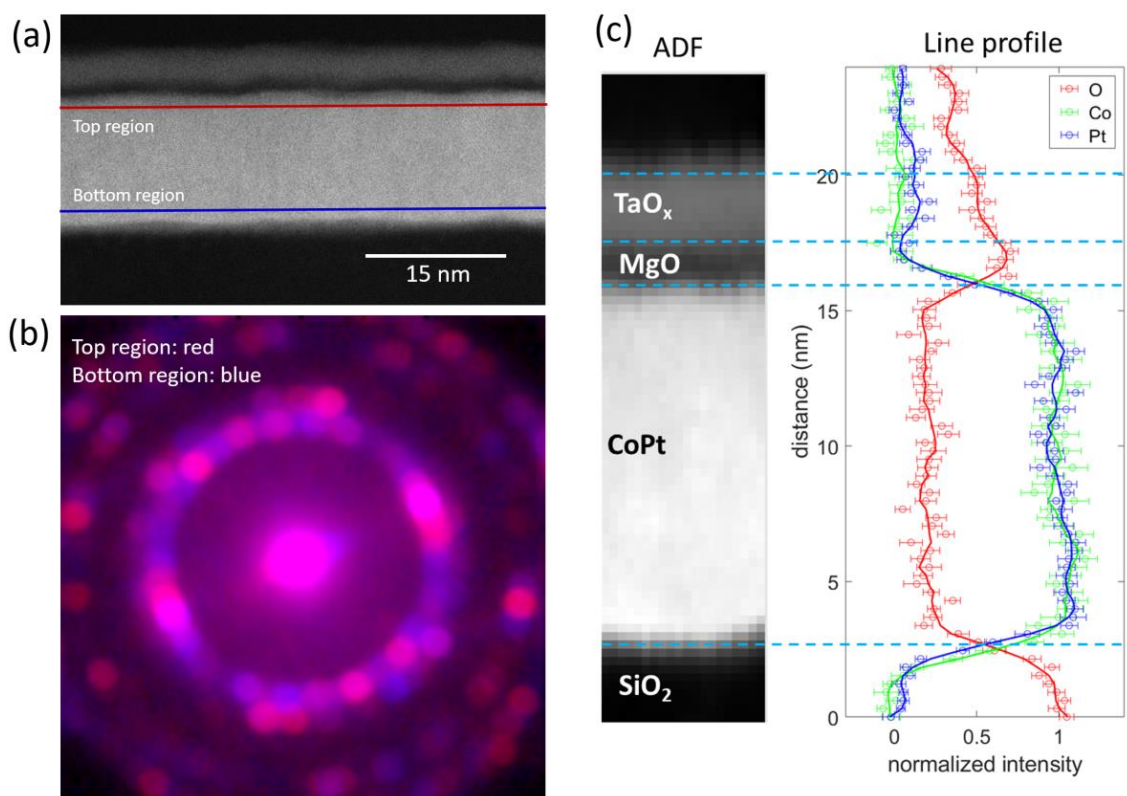

Figure S1. Structural characterizations of the sample SiO₂/CoPt 16 nm/MgO 2 nm/Ta 1.5 nm. (a) Annular Dark Field scanning transmission electron microscopy image, suggesting good compositional homogeneity in the CoPt layer. (b) Electron diffraction patterns averaged from the top region (red) and bottom region (blue) indicated by the two lines in (a), indicating chemically disordered face-centered cubic structures. The vertical strain gradient suggested by the x-ray diffraction measurement in the main text cannot be directly identified within the resolution of the electron diffraction since the red and blue patterns appear to overlap. (c) Depth profile of the electron energy loss spectrum intensities of Co, Pt, and O, revealing the absence of a composition gradient and oxidization in the CoPt. Reprinted with permission from ref. 18.

## Section 2. Self-demagnetizing factor

As theoretically predicted (see ref. 30 in the main text), a magnetic layer can be reasonably approximated as a uniform magnet with average dipole field and other dynamic variables in the analysis of the DMI in the BLS measurement for $P < 0.5$. Here, $P = 1-[1-exp(-|k|t_{FM})]/|k|t_{FM}$ is the self-demagnetization factor describing the average dipole field generated by the in-plane component of the uniform magnon mode, $k$ is wave vector, and $t_{FM}$ is the thickness of the ferromagnetic layer.

In the present work, $kt_{FM} \lesssim 0.48$, which corresponds to $P \lesssim 0.2$. Please see Fig. S2 for calculated $P$ values.

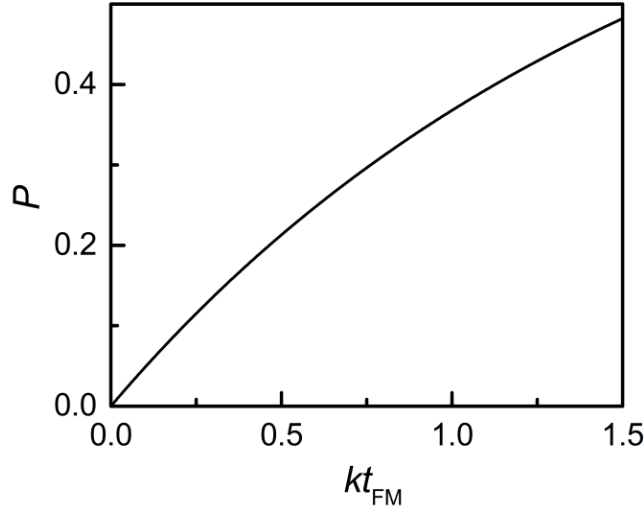

Fig. S2. Estimated $P$ plotted as a function of $kt_{FM}$.

## Section 3. A model of generation of bulk DMI by a strain gradient

While there is no detectable strain gradient in the CoPt samples in this study, we propose here that a strain gradient resulting from an accelerating strain relaxation along film normal could potentially break the inversion symmetry that can allow bulk DMI and other bulk symmetry-breaking effects to occur and vary with the thickness.

First, just like the CoPt films in this work, a magnetic thin films can be easily strained during the sputtering deposition due to the thermal stress and/or substrate stress[48-52]. When the film is thin, a strong strain survives such that strain relaxation and thus strain gradient are absent. Therefore, it is possible to have a small but non-zero threshold thickness for any asymmetry effect to occur. As the thickness of the magnetic layer ($t_{FM}$) becomes large, the strain energy will be very large and it is more favorable to relieve the strain via self-energy dislocations within the film. One can image that the rate of strain relaxation would be varying across the thickness, being slow at small first, then gradually becoming fast due to accumulation of strain energy, and finally slows down before the strain is fully relaxed due to the reduced strain energy.

For simplicity, we only discuss a model that the strain relaxation is accelerating from the bottom of the film until it is fully complete, e.g. via increasing density of dislocations (Fig. S3(a)). The strain can be characterized by the local distance of the (111) lattice plane ($d_{local}$). Here, let us assume that $d_{local}$ varies with $t_{FM}$ in a simply parabolic function, i.e. $d_{local} = -a\, t_{FM}^2 + b$, where $a$ and $b$ are constants. Then, the average distance of the (111) lattice plane ($d_{average}$) can be calculated as $-at_{FM}^2/3 + b$, which is what the x-ray diffraction can measure but $d_{average}$ varies much slower than $d_{local}$ as a function of $t_{FM}$. The slope of the strain gradient is given by $\partial d_{local}/\partial t_{FM} = -a\, t_{FM}$ so that its magnitude increases with $t_{FM}$ (Fig. S3(c)). If the local DMI, $D$, is monotonic or proportional function of the strain gradient, it would increases with the layer thickness.

If we use a more realistic strain profile function rather than the simple parabolic function to reflect the varying rate of strain relaxation (slow at small first, then gradually becoming fast, and finally slows down before the strain is fully relaxed), the strain gradient model would yield thickness dependences in $D$ and $D_s = (1/t_{FM})\int_0^{t_{FM}} D\, dt$ that are consistent well with what we have measured in Fig. 1(d) and 1(e).

We note that this strain model does not contradict the $D_s$ values being close for the composition-gradient $Co_xPt_{1-x}$ ($x = 0.75 \rightarrow 0.25$) and the composition-uniform CoPt (Fig. 2(b) in the main text). While $Co_xPt_{1-x}$ ($x = 0.75 \rightarrow 0.25$) most likely has a stronger gradient strain (the bulk A1-$Co_xPt_{1-x}$ has a cubic lattice cell in a wide composition range, with $d_{(111)}$ increasing continuously from 2.118 Å for $Co_3Pt$ to 2.225 Å for $CoPt_3$)[47], the composition gradient has also changed the strength of the spin-orbit coupling and the degree of the hybridization, both of which are very critical for the DMI. For complete, we show the x-ray diffraction results of the composition-gradient $Co_xPt_{1-x}$ ($x = 0.75 \rightarrow 0.25$), together with that of the composition-uniform CoPt, in Fig. S4.

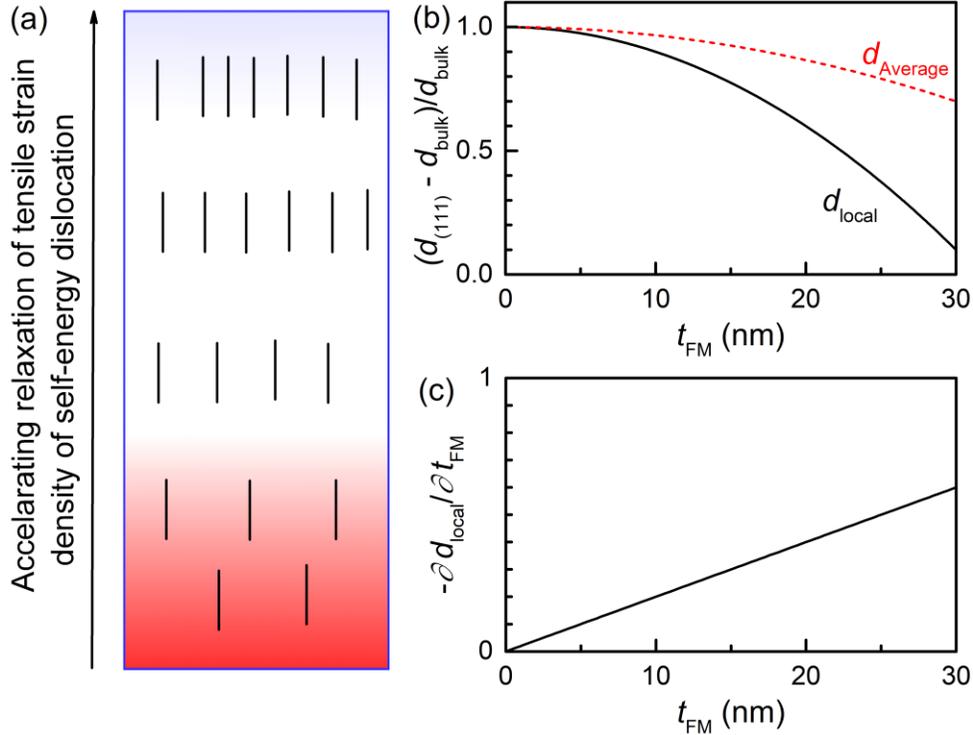

Fig. S3. (a) Schematic depict show the strain gradient induced by accelerating relaxation of the tensile strain. (b) Average ($d_{average}$) and local ($d_{local}$) spacing of the (111) lattice plane relative to the bulk value ($d_{bulk}$) and (c) $\partial d_{local}/\partial t_{FM}$ plotted as a function of $t_{FM}$. In this estimation, $d_{local} = -a\, t_{FM}^2 + b$, with $a = 0.001$ nm$^{-2}$, and $b = 1$.

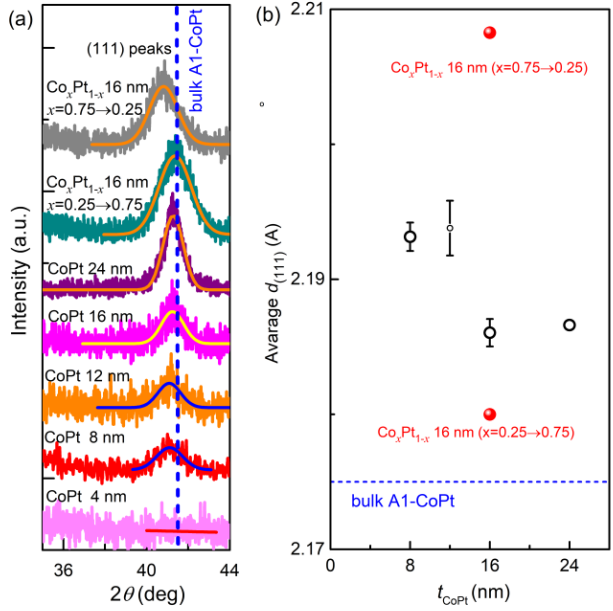

Fig. S4. (a) $\theta$-$2\theta$ x-ray diffraction patterns and (b) the average spacing of (111) lattice plane, $d_{(111)}$, for the composition-uniform CoPt films and the composition-gradient films $Co_xPt_{1-x}$ ($x = 0.25{\to}0.75$) and $Co_xPt_{1-x}$ ($x = 0.75{\to}0.25$). The solid curves in (a) represent the Gaussian fits. The blue dashed lines in (a) and (b) represent the peak position and lattice plane spacing of the bulk A1-CoPt (111), respectively.

## Section 4. Spin-torque ferromagnetic resonance

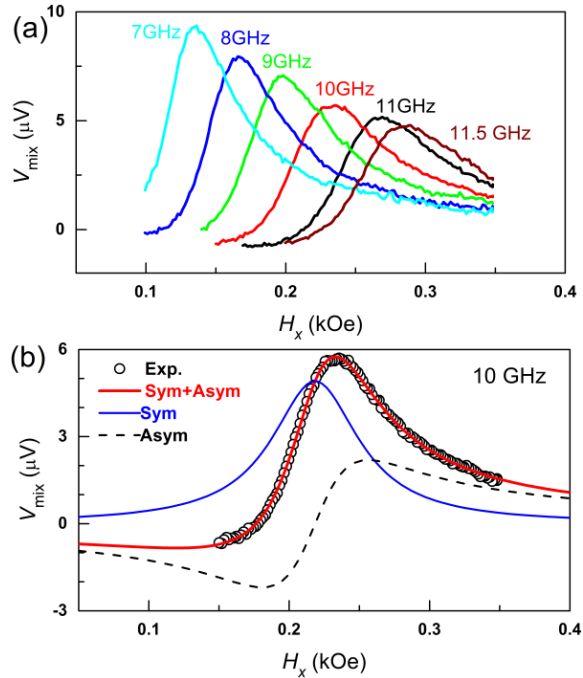

Fig. S5. Bulk spin-orbit torque-driven ferromagnetic resonance spectra for a CoPt sample: (a) 7, 8, 9, 10, 11, and 11.5 GHz and (b) 10 GHz. In (b) the blue, black, and red lines represent the symmetric component, asymmetric component, and the sum from the best fit.

4